\documentclass[review]{elsarticle}

 \usepackage{graphicx}
\usepackage{subfigure}

\usepackage{amssymb}

\journal{Computer Physics Communications}
\begin{document}

\begin{frontmatter}



\title{A Method for Incorporating General Relativity in Electromagnetic Particle-in-Cell Code}


\author{Michael Watson}

\address{Department of Physics, Fisk University, Nashville, TN 37208}

\author{Ken-Ichi Nishikawa}

\address{NSSTC/MSFC, Huntsville, AL 35805}

\begin{abstract}
An algorithm is presented that incorporates the tensor form of Maxwell's equations in a general relativistic electromagnetic particle-in-cell code. The code simplifies to Schwartzschild space-time for a non-spinning central mass. The particle advance routine uses a fourth-order Runge-Kutta algorithm
to integrate the four-velocity form of Lorentz force. The current density is calculated using the curved space-time of the metric.

\end{abstract}

\end{frontmatter}

\section{Introduction}
The algorithm described here was developed for plasma simulation in an environment around a spinning central mass. The versatility of the algorithm allows for calculations without spin. Because the algorithm uses a general metric explicitly for the description of the space-time, this algorithm can be used as a general relativistic particle-in-cell (GRPIC) code.

The basic equations used for this new code are described in Section 2. In Section 3, the numerical
scheme including discretization, field updates, and particle mover. The current deposition scheme is 
described in Section 4. Initialization and stability criteria are discussed in Section 5. Jet formation
is described as an example application of this new code in Section 6. The concluding remarks are discussed in Section 7.

\section{General relativistic particle-in-cell numerical simulation}
Our algorithm is based on a general relativistic formulation of the EMPIC algorithm \cite{bun93}. We incorporated the physical four-vectors (i.e., velocity, current and position) along with the electromagnetic field tensor to simulate the plasma particle and field dynamics.

\subsection{Formalism for GRPIC}
The equations which control the development of
the particles and fields are given by the tensor form of the
Maxwell and Newton-Lorentz equations \cite{mis73}  and the Kerr metric \cite{ker63}.

\begin{equation}
\label{Maxwell_tensor_equation:a}
F^{\alpha \beta}_{; \beta} = \frac{4\pi}{c} J^\alpha \\
\end{equation}
\begin{equation}
\label{Maxwell_tensor_equation:b}
F_{\alpha \beta;\gamma}+F_{\beta \gamma;\alpha}+F_{\gamma \alpha;\beta} =  0 \\
\end{equation}
\begin{equation}
\label{Maxwell_Newton:c} m \left(\frac{d u^\alpha}{d\tau}+
\Gamma^\alpha_{rs} \frac{dx^r}{d\tau}\frac{dx^s}{d\tau}\right) =
qF^{\alpha}_ {\beta}u^\beta
\end{equation}
where $J^\alpha$ is the four-current, $F^{\alpha \beta}$ is the contravariant electromagnetic field tensor

, $\Gamma^\alpha_{r s}$ is the Christofel symbol of the second kind for the metric, and $u^\beta$ is the four-velocity of the particle. The Latin and Greek indices have the range $(1,2,3,4)$. These equations are derived to calculate the electrodynamics of a charged particle moving in a curved space-time. The Kerr space-time in Cartesian coordinates is defined using
\begin{equation}
\label{kerr_metric}
ds^2 = -c^2d\tau^2=g_{\mu \nu} dx^\mu dx^\nu=dx^2+dy^2+dz^2-(cdt)^2+\frac{2mr^3}{r^4+a^2z^2}k^2
\end{equation}
\begin{equation}
k=\frac{r^2(xdx+ydy)+ar(xdy-ydx)+(r^2+a^2)(zdz+rdt)}{(r^2+a^2)r} 
\end{equation}
\begin{equation}
\label{radius_eqn}
r^4-(R^2-a^2)r^2-a^2z^2=0
\end{equation}
\begin{equation}
\gamma^{-2} = \left (\frac{d\tau}{dt}\right )^2=-\frac{g_{i j}v^iv^j}{c^2}
\end{equation}

where $R^2=x^2+y^2+z^2$, $m$ is the Schwarzschild mass, $ma$ the angular momentum about the $z$ axis, and $v^i=dx^i/dt$. It should be noted that in GRPIC
simulations each particle represents an ensemble of
charged particles with the finite shape over
the grids \cite{daw83,bir85,hic88}.
The underlying physics of the particle motion is governed by the tensor
form of the Newton-Lorentz equation. This form provides the equation
for  the acceleration of the particle. The acceleration is a
function of the space-time curvature defined by the metric and the
Lorentz force due to the electromagnetic field. The local field is
described by the Maxwell field tensor. The components of the tensor
are calculated using the general relativistic form of
Maxwell's equations. Using these equations the particles are
moved and the fields and currents are calculated self-consistently.

\subsection{The computational cycle}
\begin{figure}[htbp]
\centering
\includegraphics[scale=.5]{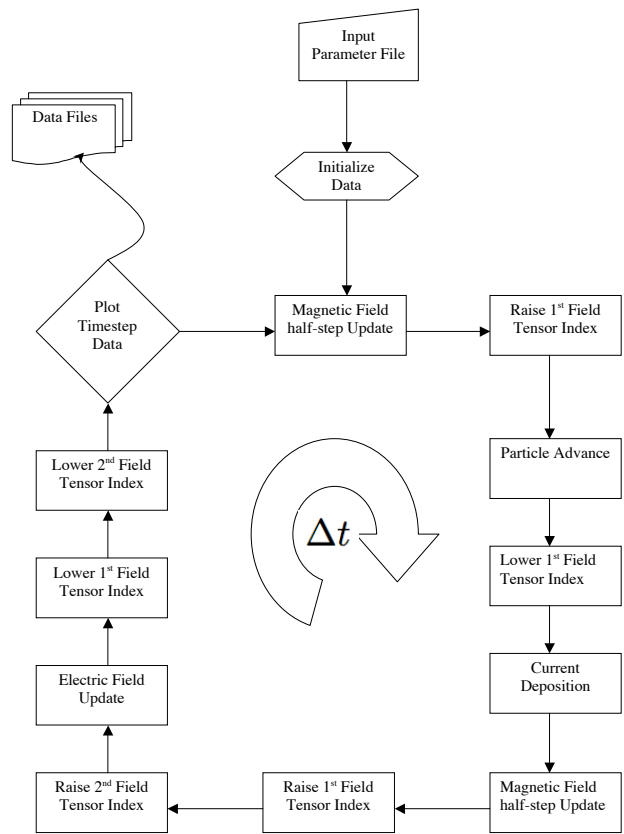}
\caption{Schematic flowchart for GRPIC simulation}
\label{flowchart}
\end{figure}

At each timestep, the algorithm solves for the fields and the particle motion. This cycle is shown in Fig \ref{flowchart}. The cycle initially starts with initial conditions on the particle positions and velocities along with the field tensor components at the grids. The particle parameters ($q, m, \mathbf{x}, \mathbf{u}$) are known at the particle location. The field tensor values are known only at discrete points on the computational cells which comprise the simulation space. The interaction between the particle parameters and the field tensor is implemented by calculating the current density on the computational grid . With the knowledge of the currents the components of the field tensor are calculated. The particles are then "pushed" using the tensor components interpolated from the grid to the particle.

The algorithm uses the Newton-Lorentz and Maxwell's equations in tensor form.

\section{Field Advance}
\label{field} 

Before and after the particles are moved, the $\mathbf{B}$ components
of the field tensor are updated in two half-steps. The   $\mathbf{E}$
components are advanced a full timestep after the particle mover and the current
calculation as shown in Figure \ref{flowchart}. The field tensor \cite{mis73} is given by $F^{4j} = -F^{j4} = E^j$,
$F^{jk} = \epsilon^{jkl}B^l$, where time component is given by $ct = x^4$. Here, $\epsilon^{jkl}$ is Levi-Civita tensor \cite{mis73}.
 
The electric and magnetic fields are components of
the Maxwell field tensor. We use the form of the contravariant tensor
\cite{mis73}. The tensor is given in the Cartesian coordinates by
\begin{equation}
F^{\alpha\beta} =
\left(
\begin{array}{cccc}
  0 & B^z  &-B^y & -E^x   \\
  -B^z & 0  & B^x& -E^y   \\
  B^y & -B^x  & 0 & -E^z  \\
  E^x & E^y & E^z & 0
\end{array}
\right)
\end{equation}
Using this form allows the quick transformation of the tensor to a covariant
or mixed form. Each of the components of the tensor are offset in space using the Yee lattice \cite{yee66} configuration.

\begin{figure}[htbp]
\centering
\includegraphics[scale=.4]{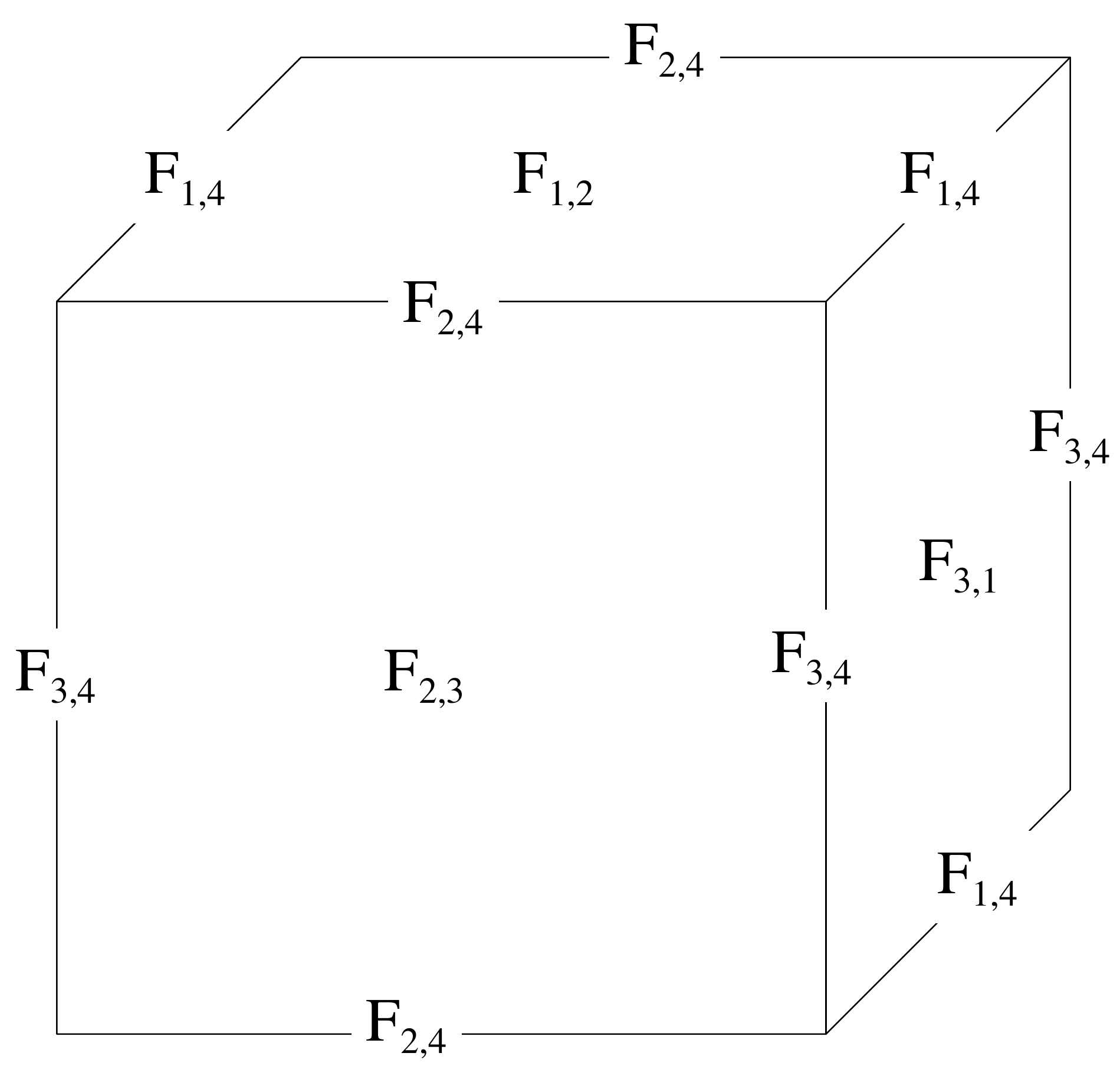}
\caption{The Maxwell tensor field components on the lattice. The components of the Maxwell field tensor are defined on the face and edge of the computational cube.}
\label{Yee_fig}
\end{figure}

An example of the covariant form of the staggered Yee lattice shown in
Figure \ref{Yee_fig} is used for the tensor update portion of the code. The same construction is used for the contravariant and mixed field tensors.

 The grid must be modified to represent the
fields in the Maxwell field tensor form. In the lattice the
components are staggered spatially and temporally shifted by
$\frac{1}{2}$.  

\subsection{Magnetic Field Update}
The magnetic shift $F_{\beta \gamma} (\mathbf{x}_i)\rightarrow F_{\beta \gamma}(\mathbf{x}_i +
\frac{\delta_{\beta i}}{2} + \frac{\delta_{\gamma i}}{2})$, where $\beta,\gamma = 1,2,3$. The magnetic field is updated by a half-timestep. We use the Yee lattice configuration and Eq. \ref{Maxwell_tensor_equation:b} to obtain a general difference equation for the magnetic field components.
\begin{equation}
F^{new}_{\alpha \beta} = F^{old}_{\alpha \beta} -
(F_{\beta \gamma;\alpha} + F_{\gamma\alpha;\beta}  )\frac{dx^\gamma}{2}
\end{equation}
where $\alpha,\beta = 1,2,3$ and $\gamma = 4$. Einstein notation does not apply.
The discretized update equation becomes
\begin{eqnarray}
F^{n+1}_{\alpha \beta}(i,j,k) & = &F^n_{\alpha \beta}(i,j,k)- \left [ \frac{F_{\beta \gamma}^n(i+\delta_{1 \alpha},j+\delta_{2 \alpha},k+\delta_{3 \alpha}) - F_{\beta \gamma}^n(i,j,k)}{\Delta x^\alpha} \right . \nonumber \\
 &  &\left . \mbox{} + \frac{F_{\gamma \alpha}^n(i+\delta_{1 \beta},j+\delta_{2 \beta},k+\delta_{3 \beta}) - F_{\gamma \alpha}^n(i,j,k)}{\Delta x^\beta}  \right ] \frac{\Delta x^\gamma}{2}
\end{eqnarray}

\subsection{Electric Field Update}
The Yee lattice is also used for the electric field update. The electric shift is
$F_{\alpha4}(\mathbf{x}_i)\rightarrow F_{\alpha 4}(\mathbf{x}_i
+\frac{\delta_{\alpha i}}{2})$ ,where $\alpha = 1,2,3,4$. The electric field
update is governed by the Eq. \ref{Maxwell_tensor_equation:a}. We write the
equation using Einstein notation. All Greek indices are in the range $(1,2,3,4)$.
Using Eq. \ref{Maxwell_tensor_equation:a} the electric field update becomes
\begin{equation}
F^{\alpha 4}_{new} = F^{\alpha 4}_{old} + \left ( \frac{4\pi}{c}J^\alpha -
\Gamma^\alpha_{\gamma \beta} F^{\gamma \beta} - \Gamma^\beta_{\gamma \alpha}
 F^{\alpha \gamma} - F^{\alpha \nu}_{,\nu}\right) dx^4
\end{equation}
The discretized update equation becomes
\begin{eqnarray}
F^{\alpha 4}_{n+1}(i,j,k) & = &F^{\alpha 4}_{n}(i,j,k)- \left [\frac{4 \pi}{c}J^\alpha_n(i,j,k) \right . \nonumber \\
 & & \mbox{} -\Gamma^\alpha_{\gamma \beta}(i,j,k) F^{\gamma \beta}_n(i,j,k)-\Gamma^\beta_{\gamma \alpha}(i,j,k) F^{\alpha \gamma}_n(i,j,k) \nonumber \\
 &  & \left . \mbox{} + \frac{F^{\alpha \nu}_n(i,j,k)) - F^{\alpha \nu}_n(i-\delta_{1 \nu},j-\delta_{2 \nu},k-\delta_{3 \nu})}{\Delta x^\nu}  \right ] \Delta x^4
\end{eqnarray}

\subsection{Integration of the equations of motion}
Because of the complexities of particle motion in curved space-time we use the fourth-order Runge-Kutta (RK4) method to integrate the four-velocity and position of each particle. This method is fourth-order accurate in time which is crucial for maintaining numerical stability and accuracy as the particles enter the region of increased curvature near the central mass. As the values of the central mass and spin increase the space-time curvature is greater.

EMPIC codes generally use the leap-frog method of Boris \cite{bor70} which is second-order. The acceleration of the particle is given by

\begin{equation}
a^\alpha = \frac{du^\alpha}{d\tau} + \Gamma^\alpha_{\mu \nu}u^\mu u^\nu = \frac{q}{m}F^\alpha_\beta u^\beta
\end{equation}
\begin{equation}
u^\alpha = \frac{d x^\alpha}{d\tau}
\end{equation}
Let
\begin{equation}
\frac{du^\alpha}{d\tau} =\tilde{a}^\alpha(\mathbf{x},\mathbf{u}) = \frac{q}{m}F^\alpha_\beta u^\beta - \Gamma^\alpha_{\mu \nu}u^\mu u^\nu
\end{equation}
and 
 \begin{equation}
\gamma_i = f(u^4_i,a^4_i) = \frac{u^4_i/c+\sqrt{(u^4_i/c)^2+4 a^4_i(\frac{\Delta t}{2c})}}{2}
\end{equation}
the steps for RK4 at timestep $n$ are:
\\
\textit{Step 1}
\begin{eqnarray*}
\tilde{u}^\alpha_1 &=& u_n^\alpha \\
x^\alpha_1 &= &x^\alpha_n \\
\gamma_1 &=& f(\tilde{u}_1^4,\tilde{a}_1^4) \\
\Delta \tau_1 &=& \frac{\Delta t}{2\gamma_1}
\end{eqnarray*}
\noindent
\textit{Step 2}
\begin{eqnarray*}
u^\alpha_2 &=& u^\alpha_n + a^\alpha_1 \Delta \tau_1 \\
\tilde{u}^\alpha_2 &=& (u_n^\alpha +  u^\alpha_2)/2 \\
x^\alpha_2 &= & x^\alpha_n + \tilde{u}^\alpha_2 \Delta \tau_1\\
\gamma_2 &=& f(\tilde{u}_2^4,\tilde{a}_2^4) \\
\Delta \tau_2 &=& \frac{\Delta t}{\gamma_2}
\end{eqnarray*}

\noindent
\textit{Step 3}
\begin{eqnarray*}
\label{ }
u^\alpha_3 &=& u^\alpha_n + a^\alpha_2 \Delta \tau_2 \\
\tilde{u}^\alpha_3 &=& (u_n^\alpha +  u^\alpha_3)/2 \\
x^\alpha_3 &= & x^\alpha_n + \tilde{u}^\alpha_3 \Delta \tau_2\\
\gamma_3 &=& f(\tilde{u}_3^4,\tilde{a}_3^4) \\
\Delta \tau_3 &=& \frac{\Delta t}{\gamma_3}
\end{eqnarray*}

\noindent
\textit{Step 4}
\begin{eqnarray*}
\label{ }
u^\alpha_4 &=& u^\alpha_n + a^\alpha_3 \Delta \tau_3 \\
\tilde{u}^\alpha_4 &=& (u_n^\alpha +  u^\alpha_4)/2 \\
x^\alpha_4 &= & x^\alpha_n + \tilde{u}^\alpha_4 \Delta \tau_3\\
\gamma_4 &=& f(\tilde{u}_4^4,\tilde{a}_4^4) \\
\Delta \tau_4 &=& \frac{\Delta t}{2\gamma_4} \\
x^\alpha_{n+1} &=& \frac{1}{6}(\tilde{u}^\alpha_1\Delta \tau_1 +  2(\tilde{u}^\alpha_2\Delta \tau_2 +\tilde{u}^\alpha_3\Delta \tau_3) +\tilde{u}^\alpha_4\Delta \tau_4)\\
u^\alpha_{n+1} &=& \frac{1}{6}(\tilde{a}^\alpha_1\Delta \tau_1 +  2(\tilde{a}^\alpha_2\Delta \tau_2 +\tilde{a}^\alpha_3\Delta \tau_3) +\tilde{a}^\alpha_4\Delta \tau_4)
\end{eqnarray*}

The general relativistic modification of the RK4 algorithm allows for the time dilation felt by each individual particle. The physics of the particle motion is locally determined by the particle position and velocity at a location in the four-dimensional space-time manifold. This insures that the total effect on the current and field tensor is due to the cumulative effect of the local space-time curvature around the particle. During the particle "push" the Christoffel symbols ($\Gamma^\alpha_{\mu \nu}$) are calculated at the current particle position. The field tensor components at the cell nodes are interpolated to the particle using a "cloud-in-cell" method \cite{bir85}. The method is modified for curved generalized space-time metrics.

\begin{figure}
\begin{center}
\includegraphics[scale=.5]{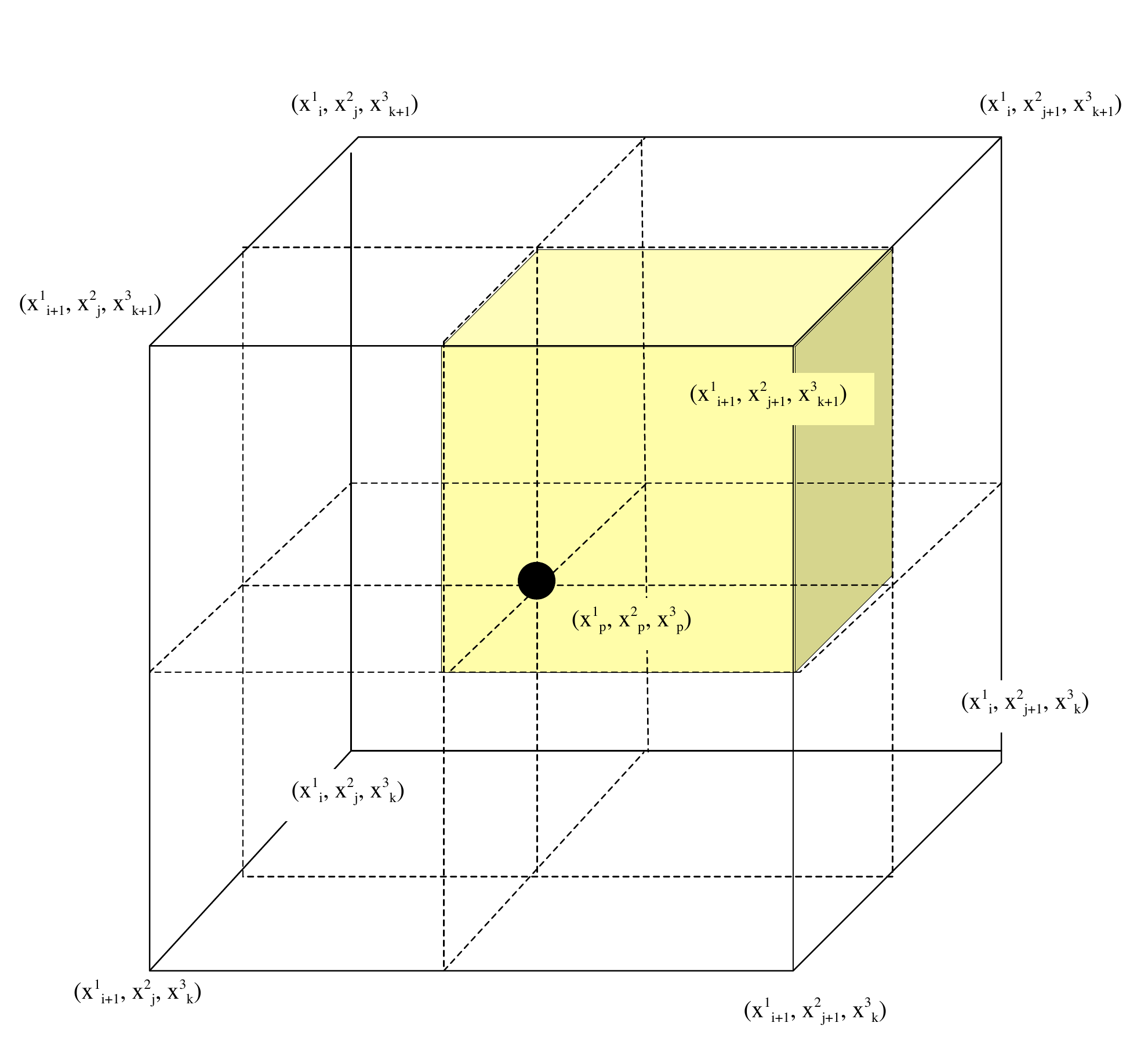}
\caption{Schematic of weighting for cell$(i,j,k)$. The shaded portion corresponds to $W_{i,j,k}$}
\label{interpolation_cube}
\end{center}
\end{figure}

The volume of each sub-cell is given by the general equations
\begin{equation}
\label{ }
 VOLUME(x^1_s,x^1_f,x^2_s,x^2_f,x^3_s,x^3_f) = \int^{x^3_f}_{x^3_s} \int^{x^2_f}_{x^2_s} \int^{x^1_f}_{x^1_s}\sqrt{-\det{(g)}}\,\,dx^1 dx^2 dx^3
 \end{equation}
  
 where $g$ is the Kerr metric. The weights for each node of the cell is given by
\begin{eqnarray*}
\mbox{CELL\underline{\,\,\,}VOLUME}(i,j,k) & = &  VOLUME(x^1_i,x^1_{i+1},x^2_j,x^2_{j+1},x^3_k,x^3_{k+1}) 
\end{eqnarray*}
\vspace*{-1.3cm}
\begin{eqnarray*}
W_{i,j,k} & = & VOLUME(x^1_p,x^1_{i+1},x^2_p,x^2_{j+1},x^3_p,x^3_{k+1})/\mbox{CELL\underline{\,\,\,}VOLUME}(i,j,k)  \\
W_{i+1,j,k} & = & VOLUME(x^1_i,x^1_p,x^2_p,x^2_{j+1},x^3_p,x^3_{k+1})/\mbox{CELL\underline{\,\,\,}VOLUME}(i,j,k) \\
W_{i,j+1,k} & = & VOLUME(x^1_p,x^1_{i+1},x^2_i,x^2_p,x^3_p,x^3_{k+1})/\mbox{CELL\underline{\,\,\,}VOLUME}(i,j,k) \\
W_{i,j,k+1} & = & VOLUME(x^1_p,x^1_{i+1},x^2_p,x^2_{j+1},x^3_k,x^3_p)/\mbox{CELL\underline{\,\,\,}VOLUME}(i,j,k) \\
W_{i+1,j+1,k} & = & VOLUME(x^1_i,x^1_p,x^2_j,x^2_p,x^3_p,x^3_{k+1})/\mbox{CELL\underline{\,\,\,}VOLUME}(i,j,k) \\
W_{i,j+1,k+1} & = & VOLUME(x^1_p,x^1_{i+1},x^2_j,x^2_p,x^3_k,x^3_p)/\mbox{CELL\underline{\,\,\,}VOLUME}(i,j,k) \\
W_{i+1,j,k+1} & = & VOLUME(x^1_i,x^1_p,x^2_p,x^2_{j+1},x^3_k,x^3_p)/\mbox{CELL\underline{\,\,\,}VOLUME}(i,j,k) \\
W_{i+1,j+1,k+1} & = & VOLUME(x^1_i,x^1_p,x^2_j,x^2_p,x^3_k,x^3_p)/\mbox{CELL\underline{\,\,\,}VOLUME}(i,j,k) 
\end{eqnarray*}

These weights can now be applied to calculate the values of the field tensor $(F^\alpha_\beta)$ at particle, p, using
\begin{eqnarray*}
F(p)^\alpha_\beta & = & F^\alpha_\beta(i,j,k)*W_{i,j,k} +  F^\alpha_\beta(i+1,j,k)*W_{i+1,j,k} +\\
 & &  F^\alpha_\beta(i,j+1,k)*W_{i,j+1,k}  +F^\alpha_\beta(i,j,k+1)*W_{i,j,k+1} + \\
 & &  F^\alpha_\beta(i+1,j+1,k)*W_{i+1,j+1,k} + F^\alpha_\beta(i+1,j,k+1)*W_{i+1,j,k+1} + \\
 &  & F^\alpha_\beta(i,j+1,k+1)*W_{i,j+1,k+1} +  F^\alpha_\beta(i+1,j+1,k+1)*W_{i+1,j+1,k+1} 
 \end{eqnarray*}

\section{Current Deposition Algorithm}

The charge conservation algorithm is based on Umeda et al \cite{ume03}. It is faster than Buneman et al \cite{bun93} and uses a straightforward coding algorithm. The current density and flux  through the faces of the cell calculations are modified to incorporate the space-time curvature. The current is calculated using the four velocity, $J=\rho \mathbf{u}=\rho \gamma \mathbf{v}$. This requires that $\gamma$ is calculated for each particle's position and velocity.

The area of the cell face given by the $x^i-x^j$-plane is
\begin{equation}
\label{ }
AREA(x^i_s,x^i_f,x^j_s,x^j_f) = \int^{x^j_f}_{x^j_s}\int^{x^i_f}_{x^i_s}\sqrt{g_{i i}g_{j j} - g_{i j}^2}\,\,dx^idx^j
\end{equation}
where $x_s$ and $x_f$ denote the start and finish locations of the nodes. The weights are given by normalizing each subarea with the total area of the face. This process is similar to the volume weighting for the particle mover. The value for the surface weights are given by
\begin{eqnarray*}
\mbox{FACE\underline{\,\,\,}AREA}(i,j) & = &  AREA(x^I_i,x^I_{i+1},x^J_j,x^J_{j+1}) \\
W_{i,j} & = & AREA(x^I_p,x^I_{i+1},x^J_p,x^J_{j+1})/\mbox{FACE\underline{\,\,\,}AREA}(i,j)  \\
W_{i+1,j} & = & AREA(x^I_i,x^1_p,x^2_p,x^J_{j+1})/\mbox{FACE\underline{\,\,\,}AREA}(i,j) \\
W_{i,j+1} & = & AREA(x^I_p,x^I_{i+1},x^J_i,x^J_p)/\mbox{FACE\underline{\,\,\,}AREA}(i,j) \\
W_{i+1,j+1} & = & AREA(x^I_i,x^I_p,x^J_j,x^J_p)/\mbox{FACE\underline{\,\,\,}AREA}(i,j) \\
\end{eqnarray*}

The current deposition in the $K$-direction at each node is then given by
\begin{eqnarray*}
J^K_{i,j} &=& FLUX^K*W_{i,j} \\
J^K_{i+1,j} &=& FLUX^K*W_{i+1,j} \\
J^K_{i,j+1} &=& FLUX^K*W_{i,j+1} \\
J^K_{i+1,j+1} &=& FLUX^K*W_{i+1,j+1} 
\end{eqnarray*}

\begin{figure}
\begin{center}
\includegraphics[scale=.4]{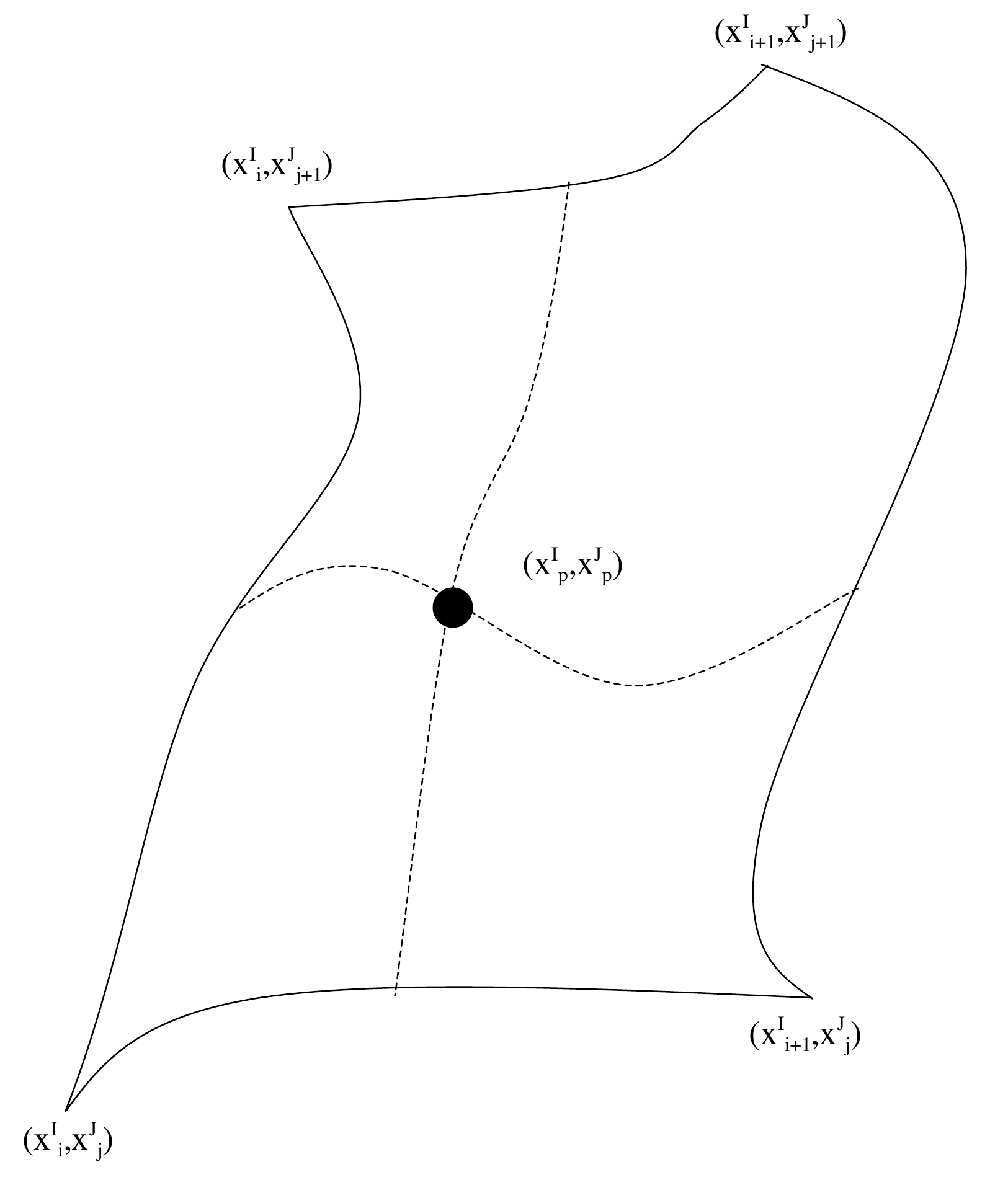}
\caption{Area schematic for generalized $x^I-x^J$ plane}
\label{ }
\end{center}
\end{figure}

\section{Initialization and Stability Criteria}
Basic criteria should be set for any PIC simulation. They are related to the plasma frequency and the Courant condition. These criteria are designed to prduce the essence of the plasma dynamics. Some of these criteria are modified and interpreted for general relativity.

\subsection{Plasma Frequency}
The plasma frequency condition parameterizes  the plasma oscillations of the system on the lowest order. The plasma frquency is given by
\begin{equation}
\omega_p \Delta t < 2
\end{equation}
\begin{equation}
\omega_p = \sqrt{4 \pi n_0 q^2/m}
\end{equation}
where $\omega_p$ is the plasma frequency, $n_0$ is the particle density, q is the charge and $m$ is the mass of the particle, $\Delta t$ is the simulation timestep. These conditions help control the numerical stablity of a PIC simulation. The parameters determined by the initial conditions of the simulation are n, q, and m.
The plasma frequency criterion is modified by the general relativistic effects of curved space-time. The partilcle density in curved space-time is transformed by $n \rightarrow \gamma n_0$. The time dilation caused by general relativity transforms $\Delta t \rightarrow \Delta t_0/\gamma = \Delta \tau$. The general relativistic plasma frequency to lowest order becomes
\begin{equation}
\omega \Delta \tau = \sqrt{4 \pi n q^2/m} = \gamma^{-1/2} (\omega_p \Delta t)<2
\end{equation}
Near the central mass, $\gamma$ will increase, therefore a given $n_0$ and $\Delta t_0$ which satisfy the plasma frequency criterion in Minkowski space will be an approximate stability condition near the central mass.
 The same arguement applies for the Courant condition. These conditions illustrate the difficulty of using classical plasma parameters as criteria for the simulation near the event horizon.

\subsection{General Relativity Considerations}

The collisionless dynamics associated with relativistic outflows can be studied using GRPIC. Specifically, the algorithm is well suited for modeling plasma environments in strong gravitational fields. The difficulty of incorporating general relativity in PIC algorithms is that the scalings are typically vastly different.

Daniel and Tajima \cite{dan97} used the PIC algorithm in a $3+1$ Schwartzchild metric. The simulation was a $1\frac{1}{2}$ dimension PIC simulation around the event horizon. The PIC algorithm is well suited for studying the acceleration mechanism around the event horizon. Also particle codes lend themselves to the incorporation of future electromagnetic radiation studies. 

General relativity uses time and length scales on the order of the Schwartzchild radius, $R_s = GM/c^2$, and time, $\tau_s = R_s/c$. PIC codes generally use scales in terms of the plasma skin depth, $\lambda_e=c/\omega_p$, and the period, $\tau=1/\omega_p$. Because of these vastly different scales a one-to-one global simulation is not feasible at this time. However, important physics can be understood by scaling the physics to dimensionless units. This allows the exploration of the dynamics of the system. For the GRPIC algorithm we use Stoney scalings \cite{ray81} to normalize the physical quantities in our simulation space. We set the standards of the system to be $\Delta x^1 = \Delta x^2 = \Delta x^3 = \Delta t = 1$, c is set to satisfy the Courant condition, and the unit charge, $\hat{q}$, is used to normalize the charge dimensions. These values give the dimensionless gravitational constant and Schwartzchild radius as
\begin{equation}
\hat{G} = \left (\frac{\Delta x c^2}{\hat{q}} \right )^2
\end{equation}
\begin{equation}
\hat{R}_S = \frac{\hat{G} M_c}{c^2}
\end{equation}
where $M_c$ is the mass of the central object.

\section{Application}
\label{application}
\begin{figure}[h] 
   \centering
   \subfigure[Initial disk and corona]{\includegraphics[scale=.10]{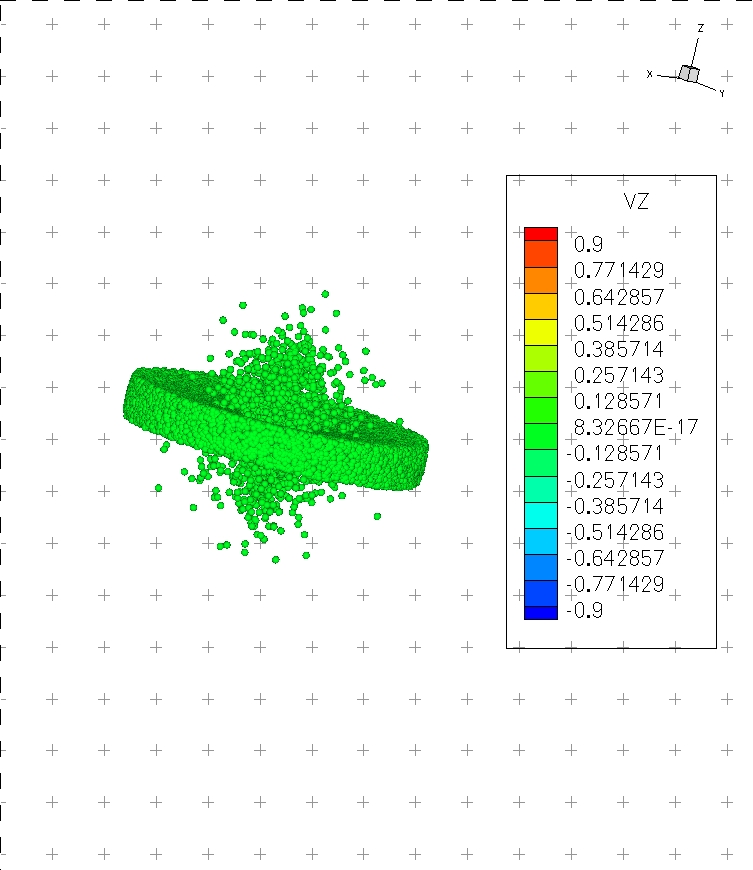}}
    \hspace{.1in}
   \subfigure[Beginning jet formation]{\includegraphics[scale = .10]{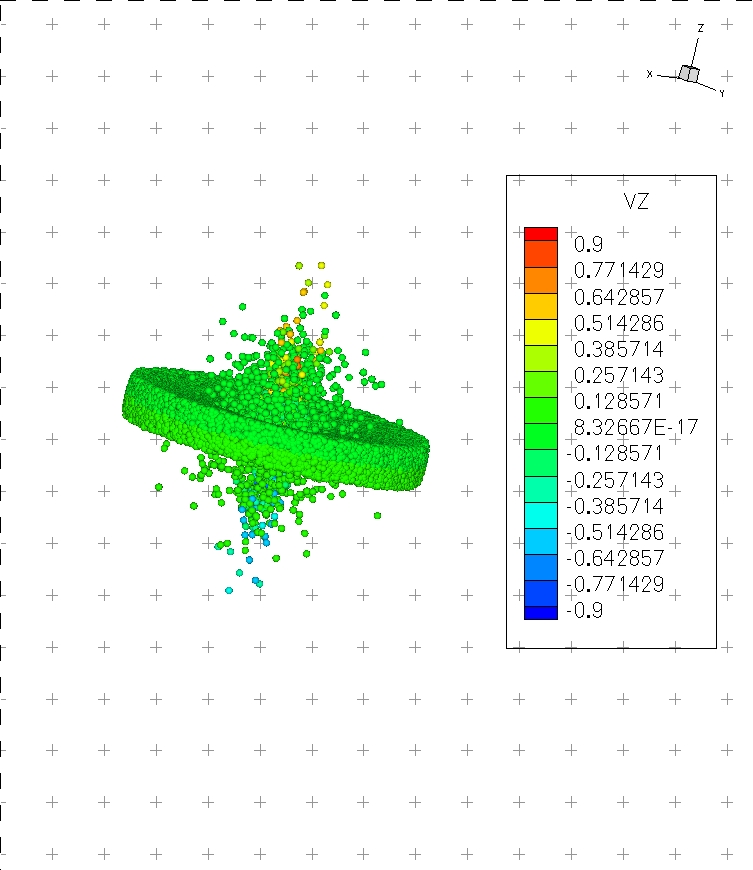}}
\hspace{.1in}
    \subfigure[Fast spiraling jet]{\includegraphics[scale = .10]{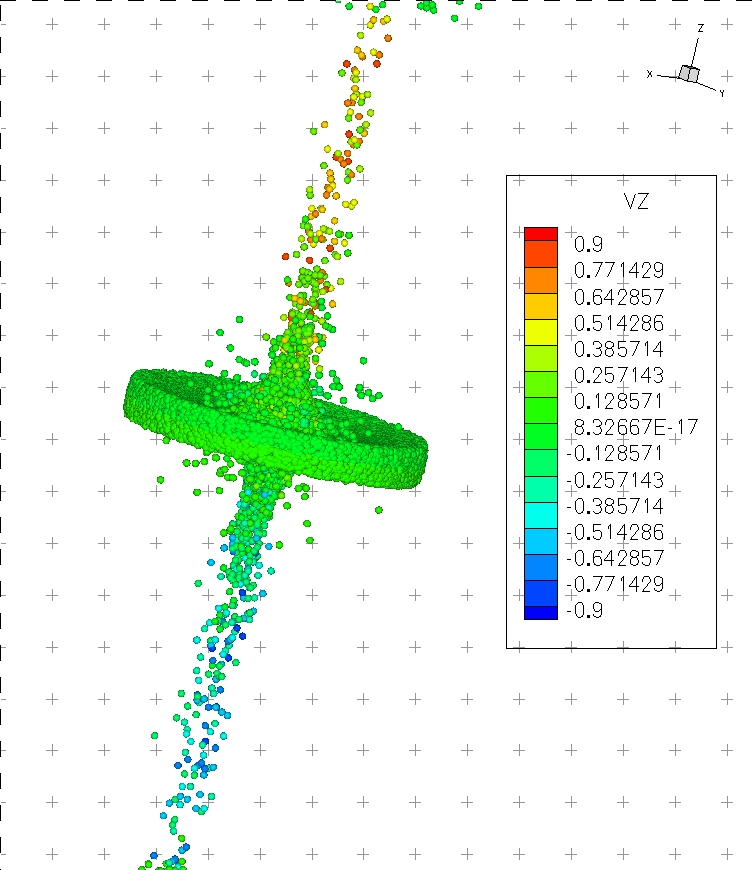}}
   \hspace{.1in}
   \subfigure[Development of slow moving outside particles surrounding faster particles at the core]{\includegraphics[scale = .10]{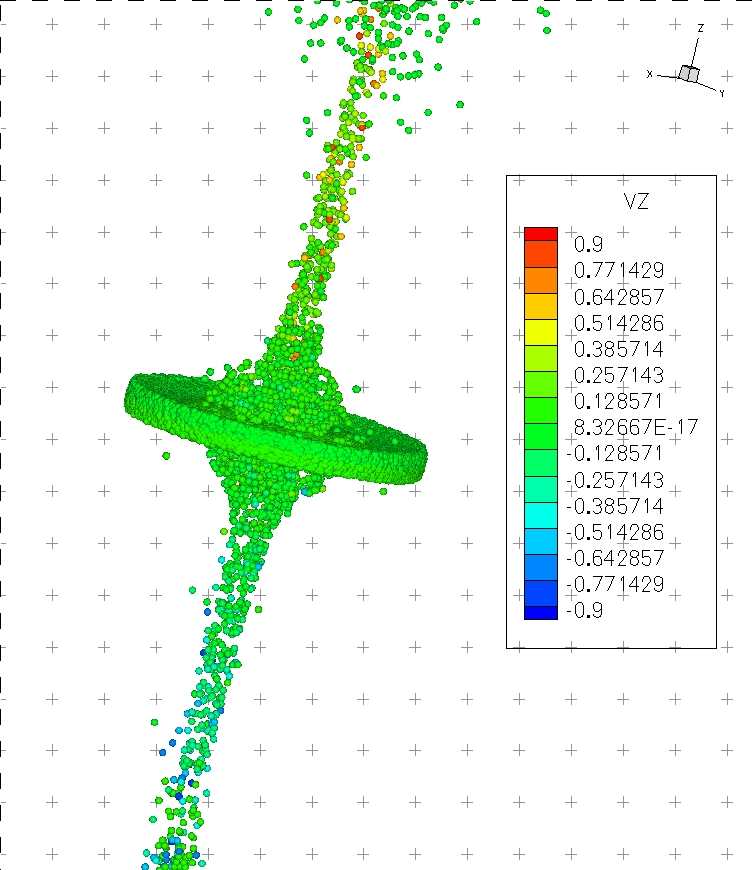}}
\caption{Jet formations are shown in the
four panels at the times $t  =$ 0, 100, 300, and 500. The jets are bipolar. Particle pairs are
   moving through the jet at different velocities. The jet has a structure
   which forms spirals around the $z$ (central)  axis. (a) The disk and corona are initialized for the simulation. (b) The initial particles are forming the jet.  (c) The jet is forming with fast spiraling particles around the central axis.  (d) The slower particles continue to move with the jet. The faster particles are at the center of the jet. }
 \label{jet}
\end{figure}

The initial disk geometry of our simulation consists of a free falling corona and a
Keplerian disk (see Figure \ref{jet}a), which is similar to Nishikawa et al \cite{nis03a,nis05b}. The central mass is located
at the center of the computational space and is co-rotating with $a=.95$. The central mass has a value
of $M_c = 300$. The Keplerian disk is located at $r>r_d\equiv
3r_s$ $|\cos{\theta}| < \delta$, where $\delta=1/8$ $r_d$ is the
disk radius. In this region the particle number is 100 times that of the corona. There are 0.8 million disk particles.
The initial orbital velocity of the particles in the disk is $v_\phi
= v_{\rm K} \equiv c\sqrt{r_{\rm S}/(2r)}$, where $v_{\rm K}$ is the
Keplerian velocity, where r is defined by Eq. \ref{radius_eqn}.
There are no disk particles initially at $r<r_d$. The initial
magnetic field is taken to be uniform in the z direction. The magnitude of the field is $30$ field simulation units. This field component is the
contravariant z component of the field. The charge-to-mass ratio of the particles is $10^{-3}$. 
Our simulation domain has a
mesh size of $64 \times 64 \times 128$. The computational space is scaled using the nodes. The grid is uniform and normalized to $\Delta x^i =
1$, where $ i = 1,2,3$. The time scaling is found by
using the Courant condition \cite{bir85}. The speed of light, c, is set to 0.05 which ensures that the particles do not move faster than the fields.
This condition also controls the growth rate of the fields and helps prevent nonphysical growth rates during the simulation.

The jet has a structure
   which forms spirals around the $z$-axis. The faster particles are at the center of the jet. The number of highly relativistic particles is much lower than the slower moving particles within the jet. The initial particles forming the jet are moving at high velocity through the infalling corona. The jet forms with fast spiraling particles around the central axis. The jet consists of a distribution of electrons and positrons. 
   
 Although the corona is falling into the central mass, slower particles are being swept up by the jet. The slower particles continue to move with the jet. The faster particles are at the center of the jet. The number of highly relativistic particles is much lower than the slower moving particles within the jet. 

\section{Conclusion}
The algorithm presented here solves a GRPIC algorithm. It incorporates a generalized metric in the numerical calculation of tensor form of Maxwell's equations and the Newton-Lorentz equation. If sufficient care is taken to satisfy the numerical stability, then this can be a very useful algorithm for using particle-in-cell codes for the simulation of astrophysical regimes where the gravity due to a central mass plays a role in the dynamics of the system.

\clearpage

\section{Acknowledgment}
M. W. was funded by NASA summer faculty fellowships (2005, 2006). This research was partially funded by a NSF subcontract PHY-0114343 with COSM (Hampton University) and the AAS Small Grant Program. M. W. is supported by UNCF Special Programs
Corporation, NASA and the NASA Administrator's Fellowship Program.

K. N. is partially supported by the National Science
Foundation award AST-0506719, and the National Aeronautic and Space
Administration award NASA-INTEG04-0000-0046, HST-AR-10966.01-A, and  NASA-06-SWIFT306-0027.

We are grateful to Y. Mizuno, C. Fendt, and Z. Kuncic for helpful discussions.

\bibliographystyle{elsarticle-num}
\bibliography{my_bibliography}

\end{document}